\documentclass[12pt]{article}

\def\gappeq{\mathrel{\rlap {\raise.5ex\hbox{$>$}}
{\lower.5ex\hbox{$\sim$}}}}

\def\lappeq{\mathrel{\rlap{\raise.5ex\hbox{$<$}}
{\lower.5ex\hbox{$\sim$}}}}

\begin{document}
\setcounter{footnote}{0}
\newcommand{\mycomm}[1]{\hfill\break
 $\phantom{a}$\kern-3.5em{\tt===$>$ \bf #1}\hfill\break}
\newcommand{\mycommA}[1]{\hfill\break
$\phantom{a}$\kern-3.5em{\tt***$>$ \bf #1}\hfill\break}
\renewcommand{\thefootnote}{\fnsymbol{footnote}}

\catcode`\@=11 
\def\lsim{\mathrel{\mathpalette\@versim<}}
\def\gsim{\mathrel{\mathpalette\@versim>}}
\def\@versim#1#2{\vcenter{\offinterlineskip
        \ialign{$\m@th#1\hfil##\hfil$\crcr#2\crcr\sim\crcr } }}
\catcode`\@=12 
\def\beq{\begin{equation}}
\def\eeq{\end{equation}}
\def\MSbar {\hbox{$\overline{\hbox{\tiny MS}}\,$}}
\def\eff{\hbox{\tiny eff}}
\def\FP{\hbox{\tiny FP}}
\def\PV{\hbox{\tiny PV}}
\def\IR{\hbox{\tiny IR}}
\def\UV{\hbox{\tiny UV}}
\def\ECH{\hbox{\tiny ECH}}
\def\APT{\hbox{\tiny APT}}
\def\QCD{\hbox{\tiny QCD}}
\def\CMW{\hbox{\tiny CMW}}
\def\pinch{\hbox{\tiny pinch}}
\def\brem{\hbox{\tiny brem}}
\def\V{\hbox{\tiny V}}
\def\BLM{\hbox{\tiny BLM}}
\def\NLO{\hbox{\tiny NLO}}
\def\res{\hbox{\tiny res}}
\def\PT{\hbox{\tiny PT}}
\def\PA{\hbox{\tiny PA}}
\def\1loop{\hbox{\tiny 1-loop}}
\def\2loop{\hbox{\tiny 2-loop}}
\def\mysim{\kern -.1667em\lower0.8ex\hbox{$\tilde{\phantom{a}}$}}
\def\a{\bar{a}}

\begin{titlepage}
\begin{flushright}
{\small CPTh/S 017.0401}\\ 
{\small CERN-TH/2001-180}\\
{\small July 2001}
\end{flushright}
\vspace{.13in}

\begin{centering}
{\Large {\bf  Conformal expansions and renormalons\footnote{Research
supported in part by the EC
program ``Training and Mobility of Researchers'', Network
``QCD and Particle Structure'', contract ERBFMRXCT980194
and the U.S. Department of Energy,
contract DE--AC03--76SF00515.}}} \\
\end{centering}

\vspace{.35in}

\begin{center}
{\bf Einan Gardi}$^{(1)}$\,\,\,and
\,\,\,{\bf Georges Grunberg}$^{(2)}$
\end{center}

\vspace{.2in}
{\small 
\noindent
$^{(1)}$\,TH Division, CERN, CH-1211 Geneva 23, Switzerland\\ 
\vspace{.1in}\\
$^{(2)}$\,Centre de Physique Th\'eorique de l'Ecole Polytechnique\footnote{CNRS
 UMR C7644}, 91128 Palaiseau Cedex, France}

\vspace{0.8in}

\noindent
{\small {\bf Abstract:}
The large-order behaviour of QCD is dominated by renormalons. On the other hand 
renormalons do not occur in conformal theories, such as the one describing the infrared
fixed-point of QCD at small $\beta_0$ (the Banks--Zaks limit). Since the fixed-point has
a perturbative  realization, all-order perturbative relations exist between the
conformal coefficients,  which are renormalon-free, and the standard perturbative
coefficients, which  contain  renormalons. Therefore, an explicit cancellation of
renormalons should occur in these relations. The absence of renormalons  in the
conformal limit can thus be seen as a constraint on the structure  of the QCD perturbative
expansion. We show that the conformal constraint is non-trivial: a generic model for the
large-order behaviour violates~it. We also analyse a specific example, based on a
renormalon-type integral over the two-loop running-coupling, where the required
cancellation  does occur.
 \vspace{.1in}\\
} 
\end{titlepage}
\newpage

In QCD, as in other quantum field theories in four dimensions, 
the running-coupling depends logarithmically on the scale. As a consequence, the perturbative
expansion is characterized by a renormalon factorial increase, which emerges from 
integrating over the running-coupling in loop diagrams~\cite{tHooft}--\cite{Beneke}.

Renormalons are believed to dominate the large-order behaviour of the series. 
In many physically interesting cases, their dominance sets in rather early~\cite{Beneke}. 
Then renormalon resummation, i.e. the summation of the specific contributions 
associated with the running-coupling to all orders, becomes important for precision calculations 
(see e.g.~\cite{Beneke}--\cite{Distribution}).

The idea that running-coupling effects can be resummed to all orders is based on an analogy with 
the Abelian theory~\cite{BLM,Disentangling}, where there is a systematic skeleton expansion.  
In the absence of the latter, the separation of running-coupling contributions
from the conformal part of the expansion is not well defined. In practice, in most applications, the 
resummation is restricted to the level of a single dressed gluon, where the (Abelian) large-$N_f$ 
calculation is sufficient to determine the renormalon contributions.   
The resummed result can then be written as
\begin{eqnarray}
\label{R_0}
R_0(Q^2)&\equiv& \int_0^{\infty} a (k^2) \,
\phi\left({k^2}/{Q^2}\right) \frac{dk^2}{k^2} \\ \nonumber
&=&a(Q^2)+ r_1^{(0)}
\beta_0a(Q^2)^2+\left(r_2^{(0)}\beta_0^2+ r_1^{(0)}\beta_1\right)
a(Q^2)^3\nonumber \\ \nonumber && +
\left(r_3^{(0)}\beta_0^3+\frac52r_2^{(0)}\beta_1\beta_0+r_1^{(0)}\beta_2
\right)a(Q^2)^4+\cdots,
\end{eqnarray}
where $k^2$ is the virtuality of the exchanged gluon,
$a(k^2)=\alpha_s(k^2)/\pi$ is the QCD running-coupling
satisfying the renormalization group equation, 
\beq
\frac{da(k^2)}{d\ln k^2}=\beta(a(k^2))=-\beta_0
{a(k^2)}^2-\beta_1{a(k^2)}^3-\beta_2{a(k^2)}^4-\cdots
\label{beta}
\eeq
and $\phi$ is an observable dependent Feynman integrand for a single gluon exchange
diagram, which is interpreted as the gluon momentum distribution function~\cite{Neu}.

Equation~(\ref{R_0}), should be viewed as the leading term in a hypothetical 
skeleton expansion~\cite{Disentangling}: $R(Q^2)=R_0(Q^2)+R_1(Q^2)+\cdots$. 
It has also proved useful~\cite{LTM,BBB,Beneke} to view
it as a model for the all-orders structure of 
the series. The Borel transform of $R_0(Q^2)$, satisfying
\beq
R_0(Q^2)=\int_0^{\infty}B(z)\,e^{-z/a(Q^2)}\,dz,
\label{Borel}
\eeq
as that of the observable $R(Q^2)$, 
has singularities (``renormalons'') on the real axis at $z=p/\beta_0$, where~$p$ are
integers, which are responsible for the factorial increase of the standard perturbative
coefficients.

On the other hand,  a conformal expansion is
totally free of renormalons. It may well contain other types of factorially increasing 
coefficients, e.g. due to the multiplicity of diagrams, but it cannot be influenced by running-coupling
effects since the coupling in the conformal theory does not run.

A simple demonstration of how conformal relations become free of renormalons is the
following:  consider~(\ref{R_0}) in the case that the coupling does not run,
\hbox{$a(k^2)=a_{\FP}$}: the integral  for~$R_0$ in~(\ref{R_0}) simply reduces to
$a_{\FP}$ (where the normalization of $\phi$ is $1$ by definition).  If instead the coupling $a(k^2)$ has an infrared
fixed-point, changing variables in (\ref{R_0}), $\epsilon \equiv k^2/Q^2$, one obtains
\beq
R_0(Q^2)=\int_0^{\infty} a \left(\epsilon Q^2\right) \,
\phi(\epsilon) \frac{d\epsilon}{\epsilon};
\label{R_0_s}
\eeq
taking the limit $Q^2\longrightarrow 0$ inside the integral one obtains again
$R_0^{\FP}=a_{\FP}$, i.e. a trivial expansion. We shall return to this example below. 
 
The fact that conformal relations are renormalon-free becomes relevant to QCD if an all-orders  
relation exists between the standard QCD perturbative expansion and a conformal expansion. 
Indeed, such a relation exists when a non-trivial infrared fixed-point is realized 
perturbatively, as occurs~\cite{Gross:1973ju}--\cite{BZ_grunberg}
for sufficiently small~$\beta_0$. Using the Banks--Zaks expansion, 
an $N_f$-independent conformal relation between two generic effective charges can be 
written~\cite{Disentangling} (see also \cite{BZ}--\cite{super}).
The coefficients in such a relation are renormalon-free. On the other hand, these coefficients are 
explicitly expressed as combinations of the standard (non-conformal) perturbative coefficients with those 
of the $\beta$ function. It immediately follows that in these combinations renormalon
factorials must conspire to cancel out. In this way, the absence of renormalons in conformal expansions translates
into a constraint on the structure of the non-conformal perturbative expansion.      

After briefly recalling the relation between conformal expansions and the standard perturbative 
expansion, we demonstrate that a generic model for 
the Borel function satisfying the expected large-order behaviour of the perturbative series is not 
necessarily consistent with the conformal limit. Next, since the conformal constraint holds by definition in~(\ref{R_0}) we analyse 
this example in detail, showing explicitly how the renormalon factorials eventually conspire to cancel out, leaving 
the conformal expansion renormalon-free.     

Suppose that the perturbative expansion of $R(Q^2)$ is given by
\beq
R(Q^2)=a(Q^2) + r_1 {a(Q^2)}^2 + r_2 {a(Q^2)}^3 +\cdots,
\label{R}
\eeq
where $a(Q^2)$ satisfies a renormalization group equation~(\ref{beta}) which has a non-trivial 
infrared fixed-point $a_{\FP}\equiv a(Q^2=0)=-\beta_0/\beta_1+{\cal O}(\beta_0)$, 
for sufficiently small $\beta_0$ (and $\beta_1<0$).
One can then write a conformal relation between
the fixed-point values of $R(Q^2=0)\equiv R_{\FP}$ and $a_{\FP}$,
\beq
R_{\FP}=a_{\FP} + c_1 {a_{\FP}}^2 + c_2 {a_{\FP}}^3 + c_3
{a_{\FP}}^4 +  \cdots,
\label{fp_relation}
\eeq 

The conformal coefficients $c_i$ can be related to $r_i$ in (\ref{R})
 through the Banks--Zaks expansion
\beq
a_{\FP}=a_0 \,+\, v_1 a_0^2 \,+\, v_2 a_0^3\, +\, v_3 a_0^4 + \cdots
\label{BZ1}
\eeq
where $a_0\equiv -\frac{\beta_0}{\left.\beta_1\right\vert_{\beta_0=0}}$,
and $v_i$ depend on the coefficients of $\beta(a)$~\cite{BZ_grunberg,CaSt,FP}.
Since $r_i$ are polynomials of order $i$ in $N_f$, one can write
\beq
r_i\equiv \sum_{k=0}^{i} r_{i,k} {a_0}^k,
\label{dij}
\eeq
yielding the following $N_f$-independent
relations~\cite{Disentangling},
\begin{eqnarray}
\label{ci}
c_1&=&r_{1,0}\\ \nonumber
c_2&=&r_{1,1} +r_{2,0} \\ \nonumber
c_3&=&-r_{1,1}v_1+r_{2,1}+r_{3,0} \\ \nonumber
c_4&=&2r_{1,1}{v_1}^2-r_{1,1}v_2-r_{2,1}v_1+r_{2,2}+r_{3,1}+r_{4,0}.
\end{eqnarray}

For simplicity we shall assume in the following that $a(Q^2)$ satisfies the two-loop renormalization group
equation. In addition, we shall ignore the $N_f$ dependence of~$\beta_1$. Under these assumptions
$a_{\FP}=a_0=-\beta_0/\beta_1$, i.e. $v_i=0$ for any
$i\geq 1$. It obviously follows that 
\beq
c_i=\sum_{k=0}^{[i/2]} r_{i-k,k}
\label{c_i}
\eeq
for any $i$, where the square brackets indicate a (truncated) integer value.
In this model then, $c_i$ is simply the sum of all the possible
$r_{j,k}$ coefficients such that $j+k=i$ and $j\geq k$.
Relaxing these restrictions is possible, but the price will be more complicated expressions. 
The conclusions would not change, provided that $\beta(a)$ itself does not contain renormalons (this is probably
true in $\overline{\rm MS}$).


Relation (\ref{c_i}) is intriguing: the $c_i$ on the l.h.s. are conformal coefficients, which must be totally 
free of renormalons, while $r_{i-k,k}$ on the r.h.s are factorially increasing because of renormalons. The only way
in which the condition that $c_i$ are renormalon-free can be realized is if explicit cancellations occur on the
r.h.s. To show that this cancellation is non-trivial, we shall now consider simple models for  
the Borel transform $B(z)$, which are consistent with the expected large-order behaviour of the series
and demonstrate that the corresponding $c_i$ 
in~(\ref{c_i}) become factorially increasing, thus violating the conformal constraint.

We begin with the simplest example corresponding to a single simple
pole in the Borel transform of the observable $R(Q^2)$:
\beq
B(z)=\frac{1}{1-(z/z_p)},
\eeq
where $z_p=p/\beta_0$ is the renormalon location.
Note that in this example we choose the renormalon residue to be
a constant, but in fact in QCD it depends on $N_f$.

The inverse Borel transform~(\ref{Borel}) yields
\beq
R(Q^2)=-z_p {\rm Ei}(1,-z_p/a)e^{-z_p/a},
\eeq
where $a=a(Q^2)$.
The perturbative series of $R(Q^2)$ 
has the following factorially increasing coefficients
\beq
r_i\,=\,i! \,\left(\frac{\beta_0}{p}\right)^i.
\label{factorial}
\eeq

Under the assumption that $\beta_1$ is $N_f$-independent, 
 the decomposition of the coefficients of (\ref{factorial}) in powers of $a_0$
according to (\ref{dij}) yields $r_{i,i}=(-\beta_1/p)^i \, i!$ and
$r_{i,j}=0$ for any $j\neq i$.
The resulting coefficients (\ref{c_i}) in this model are
therefore
\begin{eqnarray}
\label{c_i_factorial}
c_i=\left\{\begin{array}{ll}
0 &\,\,\,\,\,i \,\,\,{\rm odd}\\
(i/2)!\, \left(-\beta_1/p\right)^{i/2}&\,\,\,\,\,i \,\,\,{\rm even}\\
\end{array}
\right. .
\end{eqnarray}
Thus, the would-be ``conformal coefficients'' do diverge factorially.  In some sense
the factorial divergence is slowed down: $c_i$ contains just $(i/2)!$ rather
than $i!$.  Consequently we define $u=i/2$ and write the Banks--Zaks
expansion as:
\beq
R_{\FP}\,=\, a_0 \, \sum_{u=0}^{\infty} u! \,
\left(\frac{-\beta_1}{p}\right)^u {a_0}^{2u}
=a_0 \, \sum_{u=0}^{\infty} u! (-\delta)^{-u},
\label{R_simple_pole}
\eeq
where
\beq
\delta\equiv \frac{ p \beta_1}{\beta_0^2}.
\label{delta}
\eeq
We found that in the simple Borel pole example the factorial divergence of the
perturbative series does enter the ``conformal coefficients''. 
The conformal constraint is therefore  explicitly violated.

However, this example is not completely self-consistent: on one hand it was assumed
that $a(Q^2)$ runs according to the two-loop $\beta$ function (it has a
fixed-point), but on the other hand we used the large-$N_f$ (i.e. one-loop $\beta$
function) form of the Borel singularity, namely a simple pole.
It is known that a non-vanishing two-loop coefficient in
the $\beta$ function modifies the Borel singularity to be a branch
point.  For instance, for the leading infrared renormalon associated
with the gluon condensate ($p=2$) one has the following singularity structure
in the Borel plane~\cite{Mueller}
\beq
B(z)=\frac{1}{[1-(z/z_p)]^{1+\delta}},
\label{cut}
\eeq
where $\delta$ is defined in (\ref{delta}).
The corresponding perturbative coefficients are
\beq
r_i=\frac{\Gamma(1+\delta+i)}{\Gamma(1+\delta)}
\left(\frac{\beta_0}{p}\right)^i.
\label{di_Gamma}
\eeq
The large-order behaviour is
\beq
r_i^{as}=\frac{1}{\Gamma(1+\delta)} \,i!\,
\left(\frac{\beta_0}{p}\right)^i\,i^{\delta},
\label{di_Gamma_aprox}
\eeq
which is different from~(\ref{factorial}).

As opposed to the previous example, the
$r_i$ are not polynomials in
$\beta_0$, so starting with (\ref{di_Gamma}) we cannot obtain the
form (\ref{dij}).
To see this, let us examine the expansion of the $\Gamma$ function in $r_i$
\beq
f_i(\delta)\equiv
\frac{\Gamma(1+\delta+i)}{\Gamma(1+\delta)}=(\delta+i)(\delta+i-1)(\delta+i-2)
\dots (\delta+1).
\eeq
It is clear that $f_i(\delta)$ can be written as a sum
\beq
f_i(\delta)=\sum_{k=0}^{i} f_k^{(i)}\delta^k,
\label{f_i_delta}
\eeq
where $f_k^{(i)}$ are numbers.
Since $\delta$ is proportional to $1/{\beta_0}^2$, $f_i(\delta)$
contains all the even powers of $1/\beta_0$ from $0$ up to $2i$.
The additional positive power of $\beta_0$ in (\ref{di_Gamma}) finally
leads to having half of the terms with positive power of $\beta_0$
and half with negative powers.  The latter correspond to non-polynomial
functions of $N_f$, which cannot be obtained in a perturbative
calculation.  This suggests that the current example is unrealistic.

The actual QCD situation, where the coefficients are polynomials in $\beta_0$ 
and behave asymptotically as~(\ref{di_Gamma_aprox}), can be imitated by 
truncating the negative powers of~$\beta_0$ in~(\ref{di_Gamma}).
We verified explicitly (see below) that this truncation does not alter the 
eventual large-order behaviour of $r_i$. Note that there is some ambiguity in
the truncation: one can, in principle, truncate (\ref{f_i_delta}) at
different $k$ values and still obtain the same asymptotic behaviour.
We shall choose the most natural possibility: truncate just the
terms that lead to negative powers of $\beta_0$.

In order to proceed we should find the coefficients $f_k^{(i)}$.
This can be done by writing a recursion relation using the property
$f_i(\delta)=(\delta+i)f_{i-1}(\delta)$.
This condition is equivalent to the following
\beq
    f_k^{(i)} =\left\{\begin{array}{lllr}
   &   1                         &\,\,\,\,\,\,\,\,\,\,  &     k=i  \\
   &   f_{k-1}^{(i-1)}+if_k^{(i-1)}  &  &    0<k<i  \\
   &   i!                        &  &     k=0
\end{array}\right.
\eeq
It is straightforward to use these recursion relations to obtain $f_k^{(i)}$ to
arbitrarily high order.

After truncating the terms with negative powers of $\beta_0$, the
coefficients (\ref{di_Gamma}) become
\beq
\tilde{r}_i= \left(\frac{\beta_0}{p}\right)^i\,
\sum_{k=0}^{[\frac{i}{2}]} f_k^{(i)} \,\delta^k.
\label{di_Gamma_trunc}
\eeq
In order to verify that the truncation of the high powers of
$\delta$ does not affect the large-order behaviour of the series, we
calculated the ratio
\beq
\tilde{r}_i/{r_i}
=\left[\sum_{k=0}^{[i/2]} f_k^{(i)}\delta^k\right]\left/
\left[\sum_{k=0}^{i} f_k^{(i)}\delta^k\right]\right.
\eeq
for various values of $\delta$, as a function of the order $i$. It
turns out that this ratio approaches $1$ fast, indicating a common
asymptotic behaviour. For instance, for
\hbox{$\delta=462/625$}, corresponding to eq.~(\ref{delta}) with $p=1$ and
the values of $\beta_0$ and $\beta_1$ in QCD with $N_f=4$,
we find $\tilde{r}_i/r_i\simeq 0.995$ for $i=8$.

Next we write the decomposition of $\tilde{r}_i$ as a
polynomial in $a_0$ according to (\ref{dij}),
\beq
\tilde{r}_{i,j}=f^{(i)}_{\frac{i-j}{2}} (-1)^{j}
\left(\frac{\beta_1}{p}\right)^{\frac{i+j}{2}},
\eeq
where $j$ is odd for odd $i$ and even for even $i$ (as always, $j\leq i$).
Finally, the ``conformal coefficients'' (\ref{c_i}) in this example are
\beq
\tilde{c}_{2u}=\sum_{j=0}^{u} \tilde{r}_{2u-j,j}=\left[
\sum_{k=u}^{2u}f_{k-u}^{(k)}(-1)^{k}\right] \left(\frac{\beta_1}{p}\right)^u
\label{c_2u}
\eeq
and the expansion is
\beq
R_{\FP}\,=a_0 \, \sum_{u=0}^{\infty} \left[
\sum_{k=u}^{2u}f_{k-u}^{(k)}(-1)^{u+k}\right] (-\delta)^{-u}.
\label{R_cut}
\eeq
The square brackets should be compared with $u!$ in eq.~(\ref{R_simple_pole}), 
characterizing the simple Borel pole example.
It turns out that the $\tilde{c}_{2u}$ increase {\em faster} than
$u!$, but slower than $(2u)!$. Thus the factorial behaviour of the
``conformal coefficients'' persists also in this example. Again, 
the conformal constraint is violated.

Another possible approach to the analysis of the Borel cut example (\ref{cut})
is the following: consider the large-order behaviour of the coefficients
in eq.~(\ref{di_Gamma_aprox}).
Let us now ignore the $1/\Gamma(1+\delta)$ factor, which does not depend on $i$ and 
can be absorbed into the residue of the renormalon and expand
$i^\delta\sim \,\exp(\delta\ln(i))=1+\delta\ln(i)+\frac{1}{2}\delta^2
\ln^2(i)+\cdots$.
Again we find that high powers of $\delta$ lead to non-polynomial
dependence of the coefficients. As before we truncate these terms and
write an approximation to $r_i^{as}$:
\beq
\bar{r}_i\,=\,i!\,\left(\frac{\beta_0}{p}\right)^i
\sum_{k=0}^{[\frac{i}{2}]}\frac{1}{k!}
\ln^k(i)\ \delta^k.
\label{di_Gamma_bar}
\eeq
We checked numerically that the ratio $\bar{r}_i/r_i^{as}$
approaches $1$ as $i$
increases, so the asymptotic behaviour is not altered by this truncation.
Note that the powers of $\ln (i)$ 
in (\ref{di_Gamma_bar}) can be understood diagrammatically, 
as explained in \cite{Zakharov}.

We proceed and write
\beq
\bar{r}_{i,j}\,=\,\frac{i!}{\frac{i-j}{2}!}(-1)^j
\left(\ln i\right)^{\frac{i-j}{2}}
\left(\frac{\beta_1}{p}\right)^{\frac{i+j}{2}} ,
\label{dij_bar}
\eeq
and finally, using (\ref{c_i}), the ``conformal coefficients'' are
\beq
\bar{c}_{2u}=\sum_{j=0}^{u} \bar{r}_{2u-j,j}=\left[
\sum_{k=u}^{2u}\frac{k!}{(k-u)!}(-1)^{k}(\ln k)^{k-u}\right]
\left(\frac{\beta_1}{p}\right)^u.
\label{c_bar_2u}
\eeq
The large-order behaviour of $\bar{c}_{2u}$ turns out to be again between
$u!$ and $2u!$. In fact, the two ways we used to construct
the coefficients in this example lead to roughly the same asymptotic
behaviour of the ``conformal coefficients'': the ratio between
$\bar{c}_{2u}$ and $\tilde{c}_{2u}$
approaches a geometrical series at large orders.
The reason for this is simply the fact that in both examples the
dominant term in the sum is the one coming from the highest power of
the coupling ($\tilde{r}_{2u,0}$ in (\ref{c_2u}) and
$\bar{r}_{2u,0}$ in (\ref{c_bar_2u})).
In fact, the contributions
to the ``conformal coefficients'' from increasing orders in the coupling
are monotonically increasing in both cases. We stress, however, that
the decomposition of $r_i$ into polynomials in $\beta_0$ ($r_{i,j}$)
is not at all similar in the two cases.


We saw that, in general, consistency with the large-order behaviour of the perturbative series
does not guarantee consistency with the conformal limit. In the above examples the renormalon factorials
in $r_{i,j}$ do not cancel out in the sum (\ref{c_i}) but rather induce a non-physical factorial increase 
in the ``conformal coefficients''. Apparently, the conformal constraint is non-trivial. 

The examples above also teach us that the large-order behaviour of the
perturbative coefficients
$r_i$ (or the nature of the Borel
singularity) by itself does not uniquely
determine the decomposition of $r_i$ into powers of $\beta_0$:
several choices of $r_{i,j}$ can fit. Indeed, we shall see below that 
the cancellation of renormalons in (\ref{c_i}) crucially depends on 
having an ``appropriate'' decomposition of~$r_i$. 
Since the model (\ref{R_0}) is, by definition, consistent with the conformal constraint,
the factorials should cancel out in (\ref{c_i}) and the model should give an example
of an ``appropriate'' decomposition.  

In order to analyse the conformal coefficients in (\ref{R_0}) we restrict ourselves 
to the contribution to $R_0(Q^2)$ from small $k^2$, which is the origin of
infrared renormalons, and expand the momentum distribution function
\beq
\phi\left({k^2}/{Q^2}\right)=\sum_n
\left(\frac{k^2}{Q^2}\right)^n \, \Phi_n,
\eeq
where $\Phi_n$ are numbers. It is sufficient to consider a specific
infrared renormalon with $n=p$, so we choose our ``observable'' as
\beq
R_0(Q^2)\equiv
\int_0^{Q^2}p\left(\frac{k^2}{Q^2}\right)^p a(k^2)\frac{dk^2}{k^2},
\label{ren}
\eeq
where the upper integration limit is set for simplicity to~$Q^2$.

It was shown in \cite{G4,DU} that if $a(k^2)$
satisfies the two-loop renormalization group equation, the Borel
representation of $R_0$ is
\beq
\label{ren_Bor}
R_0(Q^2) = \int_0^{\infty} e^{-\frac{\beta_1}{\beta_0}z}
\frac{1}{\left[1-(z/z_p)\right]^{1+\delta}} \,e^{-z/a} \,dz,
\eeq
where $a\equiv a(Q^2)$.
This integral resums all those terms in eq.~(\ref{R_0}) that
depend only on the first two coefficients of the $\beta$ function.
Note that eq.~(\ref{ren}) is well defined thanks to the infrared
fixed-point of the coupling $a(k^2)$. On the other hand, 
eq.~(\ref{ren_Bor}) is not well defined because of the
infrared renormalon, and it differs \cite{G4,DU} from eq.~(\ref{ren}) by
an ambiguous power correction.
The equality between (\ref{ren}) and (\ref{ren_Bor}) should therefore be 
understood just as an equality of the (all-order) power series
expansion of the two expressions.

To expand (\ref{ren_Bor}) we note that the Borel transform of
$R_0$ with respect to the \hbox{{\em modified coupling}~$\tilde{a}$},
\beq
\frac{1}{\tilde{a}}\equiv\frac{1}{a}+\frac{\beta_1}{\beta_0},
\eeq
coincides with the Borel transform (\ref{cut}).
Using the coefficients (\ref{di_Gamma}) we have
\beq
R_0=\sum_{i=0}^{\infty} r_i {\tilde{a}}^{i+1}=
\sum_{i=0}^{\infty}\frac{\Gamma(1+\delta+i)}{\Gamma(1+\delta)}
\left(\frac{\beta_0}{p}\right)^i {\tilde{a}}^{i+1}.
\label{R_sum}
\eeq
Substituting ${\tilde{a}}^{i+1}$ for
\beq
\left(\frac{a}{1+a\beta_1/\beta_0}\right)^{i+1}=
a^{i+1}\,\sum_{k=0}^{\infty}\frac{(i+k)!}{i!\, k!}\,a^k
\left(-\frac{\beta_1}{\beta_0}\right)^k
\label{a_relation}
\eeq
we obtain
\beq
R_0=
a\,\sum_{i=0}^{\infty}\sum_{k=0}^{\infty}
\frac{\Gamma(1+\delta+i)}{\Gamma(1+\delta)}\,\frac{(i+k)!}{i!\, k!}\,
\left(\frac{\beta_0}{p}\right)^i\left(-\frac{\beta_1}{\beta_0}\right)^k
a^{i+k}.
\label{R_subst}
\eeq
Defining $n=k+i$ and performing first the summation over $i$ we obtain
\beq
R_0=\sum_{n=0}^\infty r_n a^{n+1}
\eeq
with the perturbative coefficients $r_n$ given by
\beq
r_n\,=\,\left(\frac{\beta_0}{p}\right)^n\sum_{i=0}^{n}\frac{n!}{i!\, (n-i)!}\,
\frac{\Gamma(1+\delta+i)}{\Gamma(1+\delta)}\,
(-\delta)^{n-i}.\label{rn}
\eeq
Note that the model
eq.~(\ref{di_Gamma}) corresponds to keeping only the $i=n$ term in
eq.~(\ref{rn}). We now use (\ref{f_i_delta}) to expand the $\Gamma$ function and
write explicitly the dependence on $\beta_0$ (or~$a_0$).
Defining $j=2i-2k-n$, we obtain
\beq
r_n\equiv \sum_{j=-n}^{n} r_{n,j} {a_0}^j,
\label{rij}
\eeq
 with
\beq
r_{n,j}\,=\,\left(\frac{\beta_1}{p}\right)^u
\sum_{i=u}^n\,\frac{n!}{(n-i)! \,i!} (-1)^i f^{(i)}_{i-u},
\label{rnj}
\eeq
where $u\equiv \frac{n+j}{2}$.
Next, we note that for $0\leq 2u<n$, 
\beq \sum_{i=u}^{n}
\frac{n!}{(n-i)!\,i!}(-1)^{i} f^{(i)}_{i-u}\equiv 0,
\label{cancellations} 
\eeq
so the negative powers of $a_0$ are absent in eq.~(\ref{rij}).
We thus identify a major difference between this example and the examples considered
above: here the decomposition of $r_n$ into powers of $a_0$ does not lead to any
non-polynomial dependence, and truncation is not required.
Note, however, that there are non-trivial cancellations.

Finally, using (\ref{c_i}), the conformal 
coefficients\footnote{As in the previous examples $c_i$ vanishes trivially for 
odd $i$, since $i+k$ is always even in
$r_{i,k}$ (eq.~(\ref{dij})), making  successive powers of $a_0$  decrease by a
factor of 2. This reflects a property of the two-loop $\beta$ function, namely the
ultraviolet log structure is such that two powers of $\beta_0$ are replaced by one
power of $\beta_1$ in the coefficients of successive powers of  $\log (k^2/Q^2)$ when
$a(k^2)$ is expanded in powers of $a(Q^2)$
(see eq.~(\ref{R_0})).} corresponding to $R_0$
are given by
\beq
c_{2u}=\sum_{j=0}^{u} r_{2u-j,j}=\sum_{k=u}^{2u} r_{k,2u-k}=
\left[\sum_{k=u}^{2u} \sum_{i=u}^{k}
\frac{k!}{(k-i)!\,i!}(-1)^{i} f^{(i)}_{i-u}\right]
\left(\frac{\beta_1}{p}\right)^u=0 ,
\label{c_r_2u}
\eeq
where the last equality was checked explicitly.
In other words, the final
result is
\beq
R_0(Q^2=0)=a_{\FP},
\eeq
in accordance with our expectations.
As explained above,
the vanishing of the conformal coefficients in this case can be
understood directly from the defining integral~$R_0$
(the polynomial $N_f$ dependence
of the $r_n$'s is also transparent from this representation). We note that contrary to
the previous examples (\ref{c_2u}) and (\ref{c_bar_2u}), in (\ref{c_r_2u}) the term
originating from the highest power of the coupling does not dominate. This is crucial
for the eventual cancellation.

In conclusion, we saw that the absence of renormalons in conformal coefficients can be
seen as a constraint on the form of the perturbative expansion: renormalon factorials
must cancel out in certain combinations.  We considered various examples for the Borel
transform, which are consistent with the same large-order behaviour, finding that this
cancellation is non-trivial. 
It would be interesting to find a concrete general form of
this constraint, which still appears elusive. We note that in the two-loop example
studied here the constraint is implemented through the {\em regular} factor
$\exp\left(-\frac{\beta_1}{\beta_0}z\right)$ in eq.~(\ref{ren_Bor}), and 
therefore it involves an infinite series of sub-leading terms in the large-order 
asymptotic behaviour of the coefficients.


\begin{thebibliography}{9}

\bibitem{tHooft} 
G. 't Hooft, in ``The whys of sub-nuclear physics'', Erice 1977,
ed. A.~Zichichi, Plenum (1979), p. 94.

\bibitem{Mueller}  A.H. Mueller, {\em Nucl. Phys.} {\bf B250}  (1985) 327;
{\em Phys. Lett. } {\bf B308}, 355 (1993).

\bibitem{Zakharov} V.I. Zakharov, {\em Nucl. Phys.} {\bf B385} (1992) 452.

\bibitem{Beneke} For a recent review, see
M. Beneke,  {\it Phys. Rep.} {\bf 317} (1999) 1
[hep-ph/9807443].

\bibitem{BBB}
M. Beneke and  V.M. Braun, {\em Phys. Lett. } {\bf B348} (1995) 513
[hep-ph/9411229];
P. Ball, M. Beneke and V.M. Braun, {\em Nucl. Phys. }{\bf B452} (1995)
563 [hep-ph/9502300].

\bibitem{LTM}
C.N. Lovett-Turner and C.J. Maxwell,
{\it Nucl. Phys. }  {\bf B432} (1994) 147 [hep-ph/9505224].

\bibitem{Neu} M. Neubert, {\em Phys. Rev. } {\bf D51} (1995) 5924
[hep-ph/9502264].

\bibitem{average_thrust}
E.~Gardi and G.~Grunberg,
JHEP{\bf 11} (1999) 016 [hep-ph/9908458].

\bibitem{Distribution}
E.~Gardi and J.~Rathsman, 
``Renormalon resummation and exponentiation of soft and collinear gluon radiation in the thrust
distribution'', [hep-ph/0103217], to appear in {\it Nucl. Phys. } {\bf B}.

\bibitem{BLM} S.J. Brodsky, G.P. Lepage and P.B. Mackenzie,
{\em Phys. Rev. } {\bf D28} (1983) 228; G.P. Lepage and P.B. Mackenzie,
{\em Phys. Rev. } {\bf D48} (1993) 2250.


\bibitem{Disentangling}
S.~J.~Brodsky, E.~Gardi, G.~Grunberg and J.~Rathsman,  {\em Phys. Rev.} {\bf D63}
(2001) 094017 [hep-ph/0002065].

\bibitem{Gross:1973ju}
D.J.~Gross and F.~Wilczek,
{\em Phys. Rev. } {\bf D8} (1973) 3633.

\bibitem{Caswell:1974gg}
W.E.~Caswell,
{\em Phys. Rev. Lett. } {\bf 33} (1974) 244.

\bibitem{BZ}  T. Banks and A. Zaks, {\it Nucl. Phys. } {\bf B196}
  (1982) 189.

\bibitem{BZ_grunberg} G. Grunberg, {\it Phys. Rev. } {\bf D46} (1992)
  2228.

\bibitem{CSR}
S.~J.~Brodsky and H.~J.~Lu,
Phys.\ Rev.\ D {\bf 51} (1995) 3652
[hep-ph/9405218].

\bibitem{CaSt} P. M. Stevenson, {\em Phys.  Lett. } {\bf B331} (1994) 187 [hep-ph/9402276];
 S. A. Caveny and P. M. Stevenson, ``The Banks--Zaks
Expansion and ``Freezing'' in Perturbative QCD'', [hep-ph/9705319].


\bibitem{FP}
E. Gardi and M. Karliner,
{\it Nucl. Phys. } {\bf B529} (1,2) (1998) 383 [hep-ph/9802218].

\bibitem{super}
E. Gardi and G. Grunberg,
JHEP {\bf 03} (1999) 024
[hep-th/9810192].

\bibitem{G4} G. Grunberg,  {\em Phys. Lett. } {\bf B372} (1996) 121 [hep-ph/9512203].

\bibitem{DU} Yu.L. Dokshitzer and N.G. Uraltsev, {\em Phys. Lett. }
{\bf B380}  (1996) 141 [hep-ph/9512407].


\end{thebibliography}
\end{document}